\documentclass[10pt, doublecolumn]{IEEEtran}
\usepackage{epsfig,latexsym}
\usepackage{float}
\usepackage{indentfirst}
\usepackage{amsmath}
\usepackage{bm}
\usepackage{amssymb}
\usepackage{times}
\usepackage{enumitem}

\usepackage{algorithm}
\usepackage[noend]{algpseudocode}
\usepackage{subfigure}
\usepackage{psfrag}
\usepackage{hyperref}
\usepackage{cite}
\usepackage{lastpage}
\usepackage{fancyhdr}
\usepackage{color} 
 \usepackage{amsthm}
\usepackage{bigints}
\usepackage{array}
\usepackage{booktabs}
\newcounter{problem}
\newcounter{save@equation}
\newcounter{save@problem}
\makeatletter

\begin{document}
\title{  \vspace{-0.1em}{ Environment-Division Multiple Access: an Enabler for AI-Native Multiple Access  }}

\author{ Zhiguo Ding, \IEEEmembership{Fellow, IEEE}  \thanks{ 
  
\vspace{-1em}

Z. Ding is with the School of Electrical and Electronic Engineering, Nanyang Technological University, Singapore.  
 

  }\vspace{-1em}}
 \maketitle

In wireless communication systems, multiple access techniques are used to ensure that scarce bandwidth resources can be used to support a large number of users, where the key challenge is how to suppress multiple-access interference \cite{Rappaport}. For example, in time-division multiple access (TDMA) and frequency-division multiple access (FDMA), orthogonal time slots and frequency channels are used to ensure that users are served in an interference-free manner.  

Multiple access has long been viewed as a cornerstone of modern wireless communications due to the following two reasons. The first reason is that multiple access significantly influences transceiver designs, e.g., code-division multiple access (CDMA) transmitters need to be equipped with pseudo-noise (PN) generators, whereas orthogonal frequency-division multiple access (OFDMA) transmitters require the implementation of inverse fast Fourier transform \cite{mojobabook}. The second reason is that multiple access defines the bandwidth resources available in the network, and hence plays a fundamental role in shaping the whole network architecture. For example, in CDMA and spatial-division multiple access (SDMA) networks, it is crucial to implement load control, since there is no hard limit on the number of users a system can support. However, in OFDMA systems, load control is no longer important, and the unique feature of the OFDMA physical-layer resources, namely OFDMA subcarriers, fundamentally changes the upper-layer design. In particular, in OFDMA systems, different users can experience different channel conditions on different subcarriers, which can be used to perform scheduling, resource allocation, quality of service (QoS) provisioning, etc. 
 
Because of its profound impact on both physical-layer transmission and upper-layer networking-layer design, the development of new multiple access schemes has always been a fundamental research objective in the wireless communications community. An interesting observation for those successful examples of multiple access techniques is that they all offer new domains of bandwidth resources. For example, TDMA and OFDMA explore the time and frequency domains to combat multiple-access interference. CDMA and SDMA rely on the novel code and spatial domains to ensure that users can be served simultaneously on different codes and spatial directions, respectively \cite{Verduebook, 6542746}.  The recently developed non-orthogonal multiple access (NOMA) scheme explores the power domain, where multiple users are served at the same time slot, frequency, PN code, and spatial direction but on different transmit powers \cite{NOMAPIMRC,Nomading}. Therefore, an important criterion for a new multiple-access technique is whether a new resource domain to combat multiple-access interference can be discovered. It is worth pointing out that some recently developed novel multiple access techniques do not offer new resource domains, but rely on the combination of the existing ones. Take rate-splitting multiple access (RSMA) as an example, which can be viewed as a type of multiple-input multiple-output (MIMO) NOMA, where both the power and spatial domains are exploited \cite{9831440}. Furthermore, wavelength division multiple access (WDMA) can also be viewed as a novel evolution of SDMA, where pinching-antenna-assisted beamforming and precoding are employed to suppress co-channel interference \cite{11435305}. 

This article focuses on the recently developed environment-division multiple access (EDMA) scheme, which exploits the wireless propagation environment for realizing multiple access \cite{11488474}. The development of EDMA has been enabled by two major recent breakthroughs in wireless communications research. Firstly, wireless channel conditions have long been viewed as a non-tunable system parameter beyond the control of communication engineers. However, thanks to those recently developed flexible-antenna technologies, including reconfigurable intelligent surfaces (RISs), intelligent reflecting surfaces (IRSs), movable antennas, fluid antennas, and pinching antennas, wireless propagation environments become reconfigurable, which leads to a whole new dimension to support multi-user communications \cite{irs1,irs2,10318061,9264694,pinching_antenna2,mypa}. Second, recent research advances, including integrated sensing and communications (ISAC) and channel knowledge maps (CKMs), have made it feasible to acquire an accurate knowledge of wireless propagation environments, which were, during the past, regarded as too dynamic and complex to characterize \cite{9737357,9373011}. In this article, the key features of EDMA are first illustrated by using a few important use cases in future communication networks. Then, the interaction between artificial intelligence (AI) and EDMA is discussed, where two types of applications of AI tools to multiple access,
namely AI-assisted EDMA and AI-native EDMA, are illustrated. Finally, open problems and important directions for future research in AI-assisted EDMA are discussed.

\section{Key Features of EDMA }
\subsection{Utilizing Wireless Propagation Environments }
\begin{figure}[!]\centering \vspace{-0.2em}
    \epsfig{file=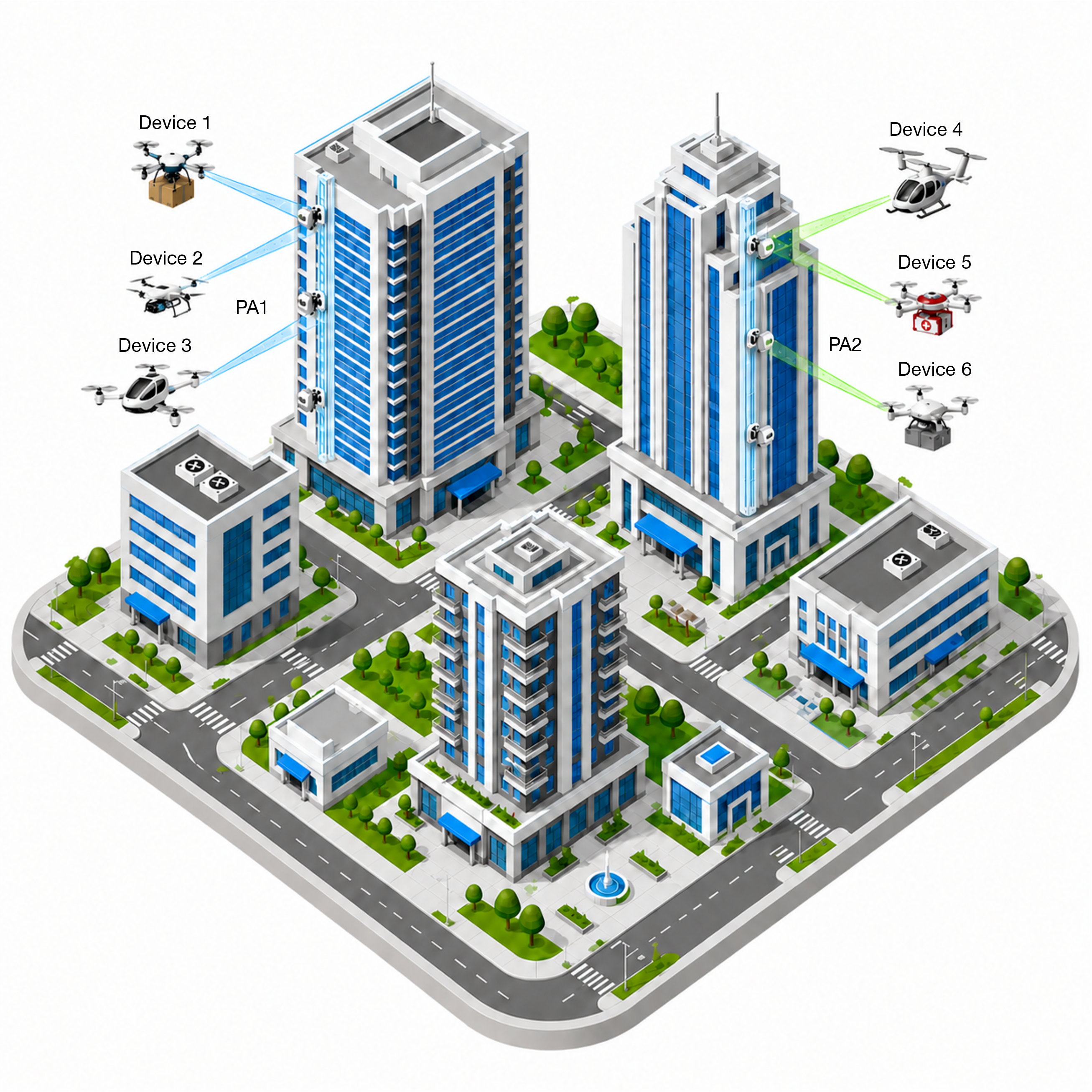, width=0.4\textwidth, clip=}\vspace{-0.5em}
\caption{Illustration for the application of EDMA to support the low-altitude economy, where ChatGPT was used to assite for the figure generation.   
  \vspace{-1em}    }\label{fig1}   \vspace{-1em} 
\end{figure}

The key idea of EDMA can be clearly illustrated by its application to the low-altitude economy (LAE), as shown in Fig. \ref{fig1}. We recall that the main challenge of LAE is how to connect a large number of LAE devices,  such as unmanned aerial vehicles (UAVs), air taxis, and delivery drones, in an ultra-reliable and low-latency manner \cite{10955337}. The existing multiple access techniques are not capable of supporting LAE. For example, the use of conventional TDMA can cause severe service delays, and employing OFDMA means that each device can have access to a small portion of available bandwidth only. Hence, they are not appropriate if LAE devices perform high-quality inspection and sensing tasks, and hence require large bandwidth with low latency. Furthermore, the complex three-dimensional LAE operating spaces also make the implementation of conventional multiple access techniques, such as SDMA, challenging.  

EDMA can exploit highly dynamic three-dimensional spaces in which LAE operate as opportunities, rather than challenges, for realizing multiple access. Take the scenario shown in Fig. \ref{fig1} as an example, where there are two high-rise buildings. Recall that line-of-sight (LoS) links are important to LAE, since LAE communication links must be reliable, low-latency, and predictable, particularly for emergency-response platforms and sensing inspectors. Therefore, pinching antennas can be deployed on the sides of the buildings to establish LoS connections to the served LAE devices. In this considered three-dimensional propagation environment, the two buildings themselves are natural LoS obstacles and can be used for interference cancellation. For example, a pinching antenna on one building, e.g.,  PA1 shown in Fig. \ref{fig1}, transmits signals to Device 3, while another pinching antenna on the other building, e.g., PA2, serves Device 6. Because of the strong LoS blockage caused by the two high-rise buildings, the LoS link between one user and its interfering antenna can be perfectly blocked. In other words, the two devices can be served on the same time and frequency channels without causing interference to each other. We further note that in EDMA, multiple-access interference can be suppressed without employing complex transmit beamforming or receive multi-user detection, which is particularly beneficial to LAE communications given the stringent energy consumption constraints of aerial devices.

\subsection{Utilizing Features of Flexible Antennas }
\begin{figure}[t]\centering \vspace{-0.2em}
    \epsfig{file=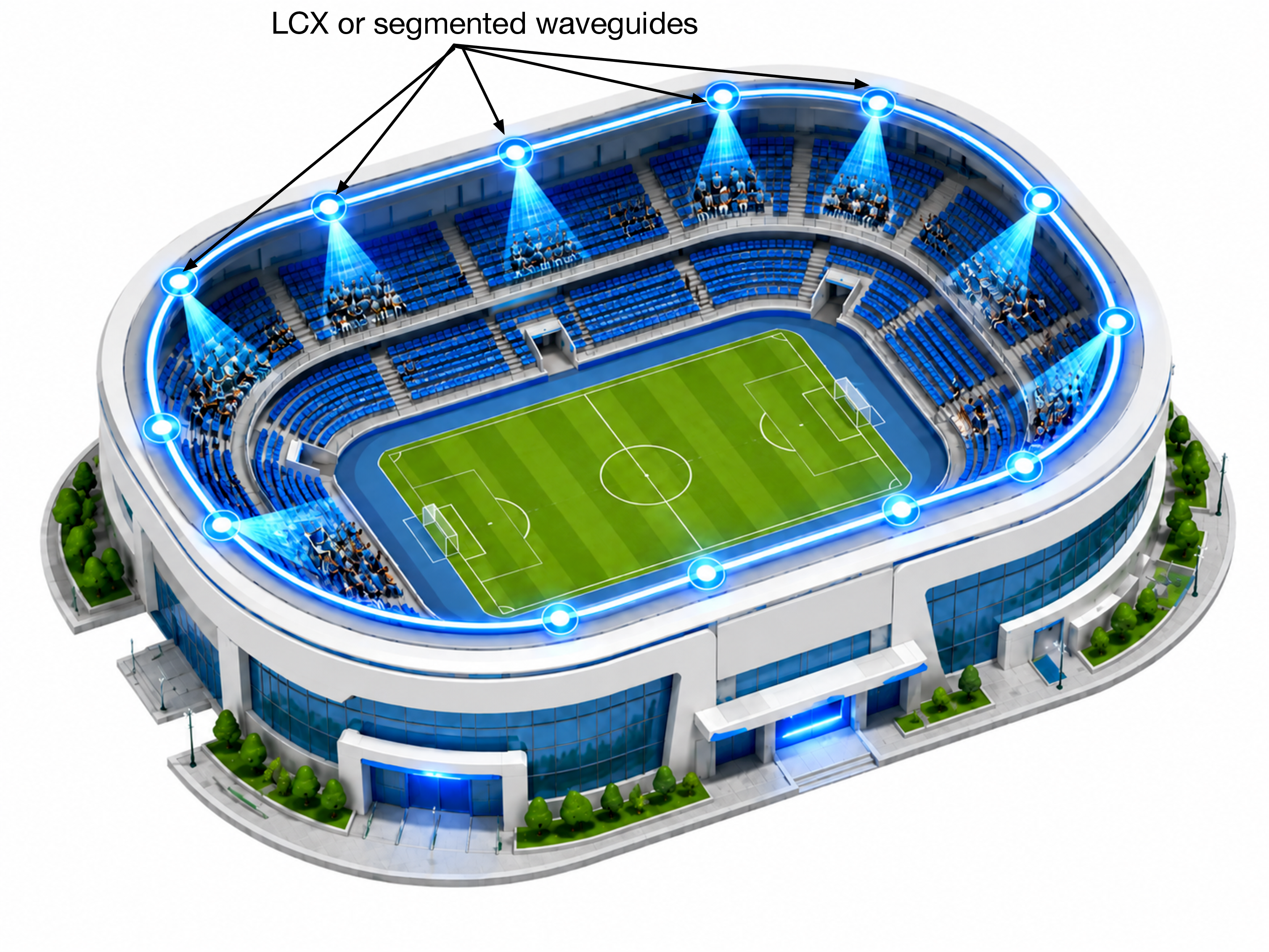, width=0.4\textwidth, clip=}\vspace{-0.5em}
\caption{Utilizing the features of flexible antennas to support massive multiple access, where ChatGPT was used to assite for the figure generation.  
  \vspace{-1em}    }\label{fig2}   
\end{figure}

Fig. \ref{fig1} illustrates a straightforward approach to passively use the key features of wireless propagation environments, such as LoS blockage, to support multiple access. The recently developed flexible antennas also offer additional degrees of freedom to facilitate the multiple access design, as illustrated in Fig. \ref{fig2}. In particular, Fig. \ref{fig2} focuses on another main challenge to design next-generation multiple access: how to support massive connectivity. Massive multiple access is particularly important for those over-crowded scenarios, such as sports stadiums, airports, etc. Most existing orthogonal multiple access (OMA) techniques are incapable of supporting massive connectivity, since the amount of bandwidth resources each user is allocated is diminishing in OMA networks. While NOMA can accommodate several times as many users as OMA networks, devices must perform complex multi-user detection to combat multiple-access interference. Conventionally, distributed antenna systems (DAS), such as cloud radio access networks (C-RAN), fog radio access networks (F-RAN), and cell-free massive MIMO, have also been proposed to support massive connectivity, but the coordination among the distributed antenna elements, such as time/phase synchronization and data sharing, can potentially lead to prohibitive system overheads \cite{CMCC,Dingpoor1311,7513863, 7827017}. 
Fig. \ref{fig2} shows an example of employing flexible antennas tailored to communication environments for supporting massive multiple access. In particular, a generalized pinching-antenna system can be preinstalled on the top of the sports stadium shown in Fig. \ref{fig2}. The key feature of generalized pinching-antenna systems, i.e., deployment flexibility and directional radio-frequency signal leakage, can be utilized to facilitate the multiple-access design and combat co-channel interference \cite{11434944}. Take the generalized pinching antennas based on leaky coaxial cables (LCX) as an example. Firstly, more antennas can be activated in the sitting areas with more users, which is facilitated by recent advances in on/off-switching-slotted LCX. Secondly, LCX's directional transmission feature can be used to suppress multiple-access interference. In particular, the coverage of such an activated generalized pinching antenna can be visualized as a conical-shaped beam, given the fact that the signal strength of a user and the antenna depends not only on their distance but also on the elevation angle associated with the user’s position relative to the LCX. Therefore, the use of generalized pinching-antenna systems essentially divides the service area, e.g., the sports stadium, into multiple non-interfering regions, where users from different regions can be served at the same time and frequency.

\subsection{Proactively Engineering Propagation Environments}

\begin{figure}[!]\centering \vspace{-0.2em}
    \epsfig{file=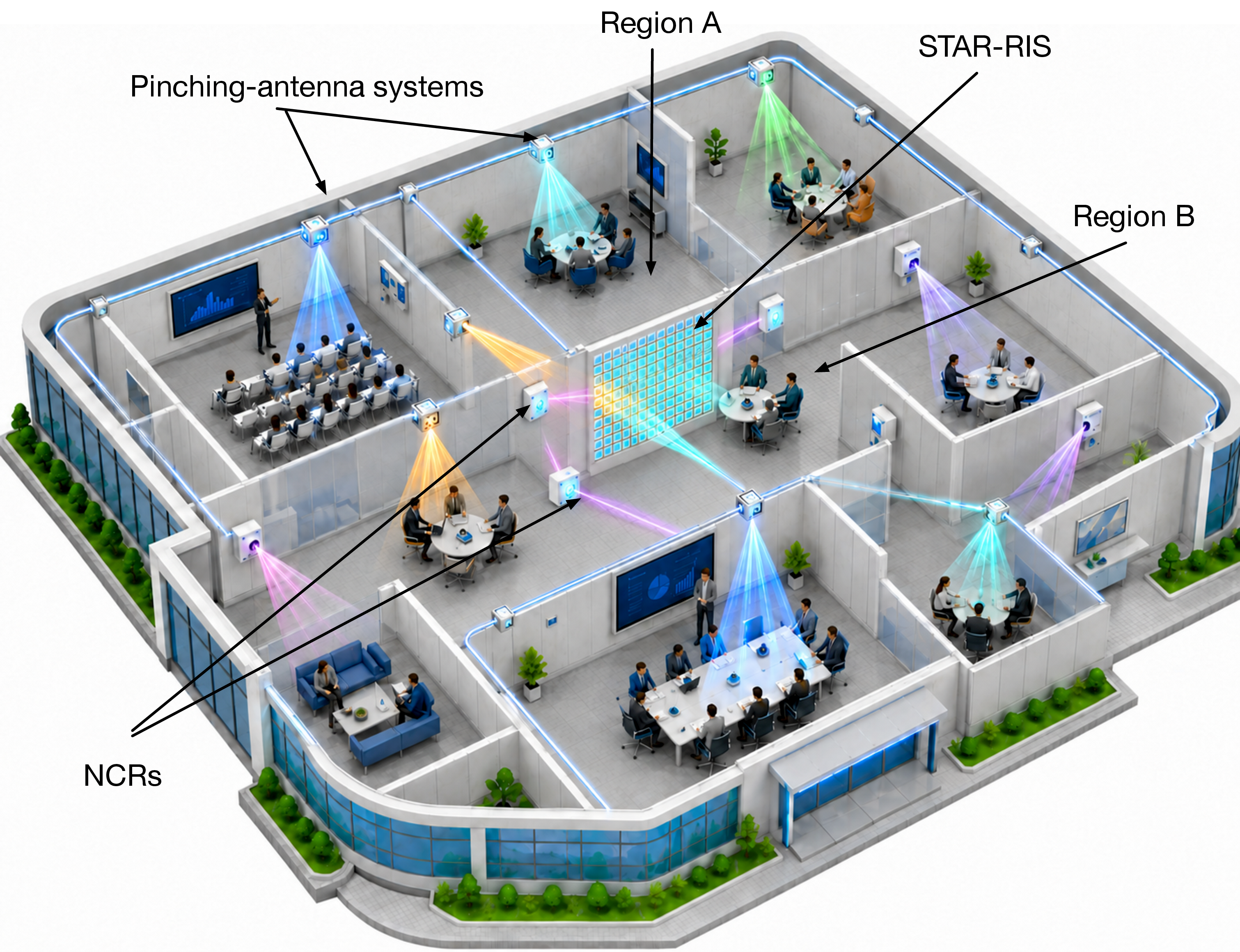, width=0.4\textwidth, clip=}\vspace{-0.5em}
\caption{Proactively engineer wireless propagation environments to support massive multiple access, where ChatGPT was used to assite for the figure generation.  
  \vspace{-1em}    }\label{fig3}   \vspace{-1em} 
\end{figure}

Wireless propagation environments can also be proactively reconfigured to support multiple access, as shown in Fig. \ref{fig3}. In particular, Fig. \ref{fig3} considers a typical indoor communication scenario, i.e., users are scattered in a conference hall which is partitioned into multiple regions. Conventionally, a single base station (or a Wi-Fi access point) is deployed in the center of the conference hall to realize the best coverage. Such a conventional approach suffers from two drawbacks. Firstly, many users in the network may be far from the base station, so their connections cannot be guaranteed. Secondly, additional users cannot be admitted if all the available bandwidth resources have been occupied. 

Assume that pinching-antenna systems are deployed in the conference hall, where preinstalled pinching antennas can be activated according to the users' instantaneous distribution. In addition, assume that the partition panel between regions A and B is a simultaneous transmitting and reflecting reconfigurable intelligent surface (STAR-RIS) which can be switched between the transmission and reflection modes \cite{9570143}. Therefore, the propagation environment in the area containing regions A and B can be dynamically reconfigured depending on the communication needs. For example, if the users in the two regions are to receive different signals, the reflective mode can be adopted by the STAR-RIS panel, such that the signals for the users on the two sides of the panel can be isolated. However, if a unicasting service is provided to the users, the transmission mode can be used, where the two regions are merged. Furthermore, assume that network-controlled repeaters (NCRs) are also deployed on the partition panels of these regions \cite{11106497}. By using the advanced capability of RISs and NCRs to reconfigure channel conditions, the wireless propagation environments can be sophisticatedly tailored to support multiple access.

\section{Towards AI-Native EDMA}
Similar to AI-native communications, there can be two stages for applying AI to the multiple-access design. The first stage is to design AI-assisted EDMA, where AI is used as a tool to tackle those specific design challenges of EDMA, such as acquiring the knowledge of propagation environments and optimizing resource allocation. The second stage is to design EDMA from the beginning with AI fully embedded into its architecture and protocols.

\subsection{AI-Assisted EDMA}
\subsubsection{AI-Assisted Environment Sensing}
Accurate knowledge of wireless propagation environments is crucial to the deployment of EDMA, since users are distinguished not only by conventional orthogonal bandwidth resources but also by their unique propagation environments. We note that the environment knowledge required by EDMA is more sophisticated than the channel condition knowledge used by conventional multiple access. In particular, in a conventional network, the primary goal of environment sensing is to estimate users' instantaneous channel coefficients, e.g., end-to-end channel gains. Such conventional sensing does not explicitly explain the cause of the estimated channel gains, i.e., how signals are propagated. The use of AI can enhance the network's knowledge of the wireless propagation environment by moving beyond conventional channel estimation toward environment understanding, which is crucial for the implementation of EDMA.

In particular, the recently developed environment-sensing approaches, such as ISAC, CKM, and digital twins, can first be used to establish a coarse understanding of dynamic propagation environments using pilot signals, beam measurements, sensing echoes, etc. Based on the data collected by these approaches, AI can be applied to infer the underlying propagation environments, including user-location information, waveguide deployment, blockage status, and mobility patterns \cite{10599123}. Take ISAC as an example, where the use of AI can transform raw sensing and communication measurements obtained from ISAC into the propagation-relevant environmental knowledge required by EDMA. In particular, the use of ISAC accomplishes the sensing for object detection and localization, and AI-assisted environment sensing can infer the communication roles of environmental objects, e.g., which object can cause LoS blockage to which user. With such AI-assisted environment sensing, it becomes feasible to transform the wireless propagation environment into a programmable and useful multiple-access resource.

\subsubsection{AI-Assisted Resource Allocation}
Resource allocation is crucial to the successful deployment of EDMA, but resource allocation in EDMA networks is more challenging than that in conventional networks, due to the following reasons. First, the use of flexible antennas is key to utilizing the properties of wireless propagation environments for the multiple access design. However, the full potential of flexible antennas can be unlocked only with sophisticated optimization, which cannot be tackled by conventional optimization tools. Take pinching antennas as an example, where the antenna placement optimization is the crown jewel for the design of pinching-antenna systems. However, the antenna placement optimization problem is highly non-convex and exhibits severe local-optimum oscillations, to which the effectiveness of conventional optimization tools, such as convex optimization and gradient-based search methods, is limited. Second, resource allocation for EDMA is cross-layer optimization by nature, where components from different communication protocol stacks, e.g., power allocation at the physical layer and user scheduling at the medium access control (MAC) layer, need to be jointly designed. This again limits the effectiveness of conventional optimization tools.

AI is an ideal tool for the aforementioned challenging optimization problems in EDMA networks \cite{8808168}. For example, in many EDMA resource allocation optimization problems, the objective functions are not explicit functions of the optimization variables, such as antenna locations and user scheduling. As a result, closed-form expressions for the derivatives of these objective functions with respect to the optimization variables are difficult to obtain, which makes the application of conventional optimization tools challenging. On the other hand, AI tools, such as unsupervised learning and reinforcement learning, do not require explicit expressions of gradients, and, hence, are ideal to be used to tackle these challenging optimization problems \cite{8714026,9448276}. Furthermore, for the aforementioned local-optimum oscillation situations, the use of AI tools can first effectively learn good initializations as well as the structure of high-quality solutions, which is important for those highly non-convex optimization problems. In addition, for the aforementioned complex cross-layer optimization problems with mixed-integer variables, AI-based optimization methods can effectively learn the properties of near-optimal solutions, and, hence, deliver high-quality decisions in a low computational complexity manner.

     \begin{figure*}[t]\centering \vspace{-0.2em}
    \epsfig{file=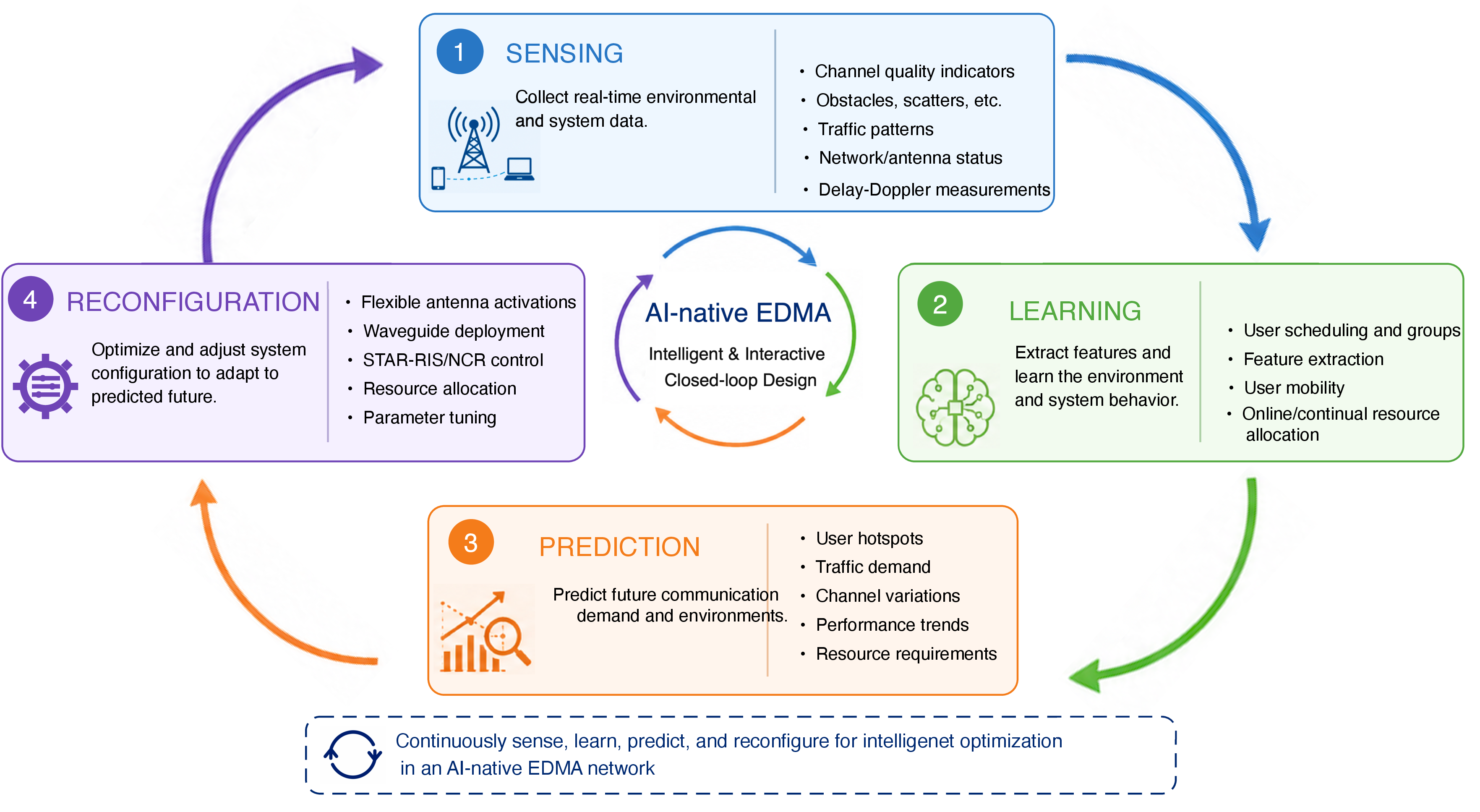, width=0.8\textwidth, clip=}\vspace{-0.5em}
\caption{Illustration for the design of AI-native EDMA.
  \vspace{-1em}    }\label{fig4}   \vspace{-1em} 
\end{figure*}

\subsection{AI-Native EDMA}
In addition to AI-assisted multiple access, where AI tools are applied to those predefined optimization problems, AI-native EDMA can be designed by fully embedding AI as an intrinsic part of the multiple-access architecture. Such a full integration of AI is crucial to the successful deployment of EDMA, since EDMA is not only to passively exploit the properties of wireless propagation environments, but also to proactively reconfigure the propagation environments in order to better support multi-user communications. Ideally, the EDMA network should be designed as a closed-loop intelligent system that can sense, learn, predict, and reconfigure the wireless propagation environment in real time, as explained in the following:
\begin{itemize}
\item Sense: AI can be used to effectively characterize the complex propagation environments by mapping those raw measurements, such as received channel quality indicators, delay-Doppler measurements, beam-sweeping results, etc, into useful information which can be used to design multiple access. For example, the echoes of sensing signals are available in a system equipped with ISAC, and they can be processed by AI tools to identify the objects in the environment, such as obstacles and reflectors, as well as to sense network user mobility.   

\item Learn: With the sensed knowledge of the propagation environment, AI can then be used to learn the impact of the environment on supporting multiple access. For example, given those obstacles sensed, the AI tools can be used to learn which users should be scheduled to utilize such obstacles and how much transmit powers they should use, e.g., users with their interference links blocked by the obstacles should be scheduled and their transmit powers can be lowered. Furthermore, the characterization of the wireless networks, such as the potential LoS paths, the spatial distribution of users, and the availability of the bandwidth for serving these users, can also be effectively learned.

\item Predict:  Unlike conventional multiple-access techniques, which are typically designed based on instantaneous channel conditions, EDMA inherently involves multi-snapshot, long-term optimization in dynamic and complex environments. Therefore, the prediction capability provided by AI is particularly crucial to EDMA. Taking the exhibition scenario illustrated in Fig. \ref{fig3} as an example, the learned user distribution in the current time slot can be utilized together with the exhibition program information to predict the future distribution of users across different exhibition-partitioned regions,  which can then be used for the following reconfiguration stage. Similarly, for the low-altitude economy scenario shown in Fig. \ref{fig1}, the learned user mobility can be used to predict the future locations of these devices, which is crucial for serving high-mobility users.

\item Reconfigure: Based on the learned and predicted knowledge, AI tools can be applied to reconfigure wireless propagation environments to better support multiple access. Take the low-altitude economy application shown in Fig. \ref{fig1} as an example. The learned/predicted knowledge about the device distribution can be used to activate pinching antennas in the proximity of the served users/devices in order to create strong LoS connections and suppress multiple-access interference. For the indoor communication scenario shown in Fig. \ref{fig3}, the STAR-RIS panel can be set to the reflective mode if users on both sides need to be served simultaneously, where NCRs can be tuned to simultaneously enhance the signal strength and suppress co-channel interferences. The predictive capability of AI can further be used to trigger manual reconfiguration of flexible-antenna systems. For example, additional waveguides, STAR-RISs, and NCRs can be deployed at hotspots that are predicted to accommodate more users in the future.
\end{itemize} 
We note that the implementation of AI-native EDMA is not a static and one-off process. Instead, the four steps, sensing, learning, prediction, and reconfiguration, need to be carried out in an intelligent and interactive manner, as shown in Fig \ref{fig4}. For example, the outcomes of learning and prediction yield specific instructions for the reconfiguration step whose impact on the propagation environments needs to be further sensed and learned. Again, take the indoor scenario shown in Fig. \ref{fig3} as an example. The network can first sense the static propagation environment, e.g., obstacles and scatterers, then learn and predict users' communication requirements, which can be used to determine where to activate pinching antennas during the reconfiguration step. Afterward, the steps for sensing, learning, and predicting need to be carried out again to make further recommendations, e.g., additional hardware for flexible-antenna systems should be deployed in overcrowded halls.

 \section{Conclusions and Future Research Directions} 
 In this article, a new type of multiple access, termed EDMA, has been introduced and its interaction with AI-native communication networks has been illustrated. In particular, the key properties of EDMA, such as utilizing the features of wireless propagation environments, integrating advanced flexible antennas, and proactively reconfiguring propagation environments, have been described. The article also illustrated two types of applications of AI tools to multiple access, namely AI-assisted EDMA and AI-native EDMA. A few promising directions for future research related to EDMA are provided as follows.

 \subsection{Identifying More Features of Propagation Environments}
 The key idea of EDMA is to exploit the features of wireless propagation environments, such as LoS blockage, large-scale path loss, and small-scale multi-path fading, for realizing the multiple access design. Therefore, identifying more features of propagation environments is crucial since it yields more degrees of freedom to design EDMA networks. To achieve this goal, some propagation properties need to be thoroughly reviewed from the multiple access perspective. For example, LoS blockage has been conventionally regarded as detrimental, but in the multiple-access context, it provides a low-complexity means to suppress co-channel interference. Furthermore, the key properties of new materials and devices should also be carefully examined and utilized. For example, the directional transmission property of LCX-based pinching antennas can be effective to create strong hotspots and also mitigate multiple-access interference. 
 
 \subsection{Enabling Dynamic Interactions between Multiple Access and Environments}
 The interaction between conventional multiple access techniques and propagation environments is limited. For example, TDMA, FDMA and CDMA are environment-free multiple access designs. While SDMA and NOMA are environment-dependent, the features of propagation environments are utilized in a passive manner. On the contrary, in EDMA networks, the interactions between multiple access and propagation environments are complex and bi-directional, where the environment features are used to support multiple access, and also the demand from multiple access leads to the reconfiguration of propagation environments. While the use of conventional convex optimization tools is important to obtain insightful understandings of EDMA, advanced AI-assisted tools are crucial to realize intelligent interactions between multiple access and wireless propagation environments.   
 
  \subsection{Application of Foundation Models}
  Recent advances in foundation models for wireless communications suggest a new paradigm for AI-native EDMA. On the one hand, the use of conventional AI models is ideal for AI-assisted EDMA, where task-specific AI models can be trained for individual functions. For example, empirical data obtained from pilot transmissions or computer-vision-based measurements can be used to develop AI-assisted channel estimation and sensing schemes. On the other hand, EDMA, assisted by a foundation model, can provide a common intelligence decision-making platform that can be adapted to diverse communication tasks. The challenge for design foundation-model-assisted EDMA is to obtain  large-scale wireless data, from which a wireless foundation model can learn reusable and generalizable knowledge. Recall that the wireless propagation environment is highly dynamic, heterogeneous, and scenario-dependent, and hence challenging to effectively learn. Furthermore, the fact that wireless data is different from natural language or image data, and is strongly affected by various factors, such as carrier frequency, user mobility, shadowing, and blockage, makes it difficult for a foundation model trained in one scenario to be generalized to other scenarios.
 
 \subsection{Private-Network Applications and Customized EDMA Designs}
 In addition to the aforementioned foundation models, another promising research direction is to focus on private networks, e.g., factories, ports, airports, campuses, hospitals, and mines, as defined for 5G systems. In these private networks, application scenarios are typically well defined, and the associated tasks and service requirements can be clearly specified in advance, where lightweight AI models can be more efficiently implemented. Take the application of EDMA to mines as an example. The private network is deployed in relatively confined and well-understood environments, where user behavior, service requirements, and the physical layout, e.g., the existence of mine tunnels and their specific locations/usage, are more predictable. This makes it possible to design lightweight AI models tailored to specific EDMA functions, such as hotspot prediction, user grouping, blockage detection, resource allocation, or the reconfiguration of flexible antennas, such as LCX in the tunnels. Compared to large-scale foundation models, such compact models require less training data, lower computational complexity, and reduced energy consumption, which ensures that EDMA equipped with task-oriented lightweight AI achieves a balanced tradeoff between realizing an intelligent solution and maintaining low implementation complexity.

\bibliographystyle{IEEEtran}
\bibliography{IEEEfull,trasfer}

\begin{thebibliography}{10}
\providecommand{\url}[1]{#1}
\csname url@samestyle\endcsname
\providecommand{\newblock}{\relax}
\providecommand{\bibinfo}[2]{#2}
\providecommand{\BIBentrySTDinterwordspacing}{\spaceskip=0pt\relax}
\providecommand{\BIBentryALTinterwordstretchfactor}{4}
\providecommand{\BIBentryALTinterwordspacing}{\spaceskip=\fontdimen2\font plus
\BIBentryALTinterwordstretchfactor\fontdimen3\font minus
  \fontdimen4\font\relax}
\providecommand{\BIBforeignlanguage}[2]{{%
\expandafter\ifx\csname l@#1\endcsname\relax
\typeout{** WARNING: IEEEtran.bst: No hyphenation pattern has been}%
\typeout{** loaded for the language `#1'. Using the pattern for}%
\typeout{** the default language instead.}%
\else
\language=\csname l@#1\endcsname
\fi
#2}}
\providecommand{\BIBdecl}{\relax}
\BIBdecl

\bibitem{Rappaport}
T.~S. Rappaport, \emph{Wireless Communications: Principles and Practice}.\hskip
  1em plus 0.5em minus 0.4em\relax Prentice Hall, New York, US, 1998.

\bibitem{mojobabook}
M.~Vaezi, Z.~Ding, and H.~V. Poor, \emph{Multiple Access Techniques for {5G}
  Wireless Networks and Beyond}.\hskip 1em plus 0.5em minus 0.4em\relax
  Springer International Publishing, 2019.

\bibitem{Verduebook}
S.~Verd\'{u}, \emph{Multiuser Detection}.\hskip 1em plus 0.5em minus
  0.4em\relax Cambridge University Press, Cambridge, UK, 1998.

\bibitem{6542746}
A.~Adhikary, J.~Nam, J.-Y. Ahn, and G.~Caire, ``Joint spatial division and
  multiplexing - the large-scale array regime,'' \emph{IEEE Trans. Inform.
  Theory}, vol.~59, no.~10, pp. 6441--6463, Oct. 2013.

\bibitem{NOMAPIMRC}
Y.~Saito, A.~Benjebbour, Y.~Kishiyama, and T.~Nakamura, ``System level
  performance evaluation of downlink non-orthogonal multiple access {(NOMA)},''
  in \emph{Proc. IEEE Int. Symposium on Personal, Indoor and Mobile Radio
  Commun.}, London, UK, Sept. 2013.

\bibitem{Nomading}
Z.~Ding, Z.~Yang, P.~Fan, and H.~V. Poor, ``On the performance of
  non-orthogonal multiple access in {5G} systems with randomly deployed
  users,'' \emph{IEEE Signal Process. Lett.}, vol.~21, no.~12, pp. 1501--1505,
  Dec. 2014.

\bibitem{9831440}
Y.~Mao, O.~Dizdar, B.~Clerckx, R.~Schober, P.~Popovski, and H.~V. Poor,
  ``Rate-splitting multiple access: Fundamentals, survey, and future research
  trends,'' \emph{IEEE Commun. Surveys Tuts.}, vol.~24, no.~4, pp. 2073--2126,
  2022.

\bibitem{11435305}
J.~Zhao, X.~Mu, K.~Cai, Y.~Zhu, and Y.~Liu, ``Waveguide division multiple
  access for pinching-antenna systems ({PASS}),'' \emph{IEEE Trans. Wireless
  Commun.}, vol.~25, pp. 13\,761--13\,775, 2026.

\bibitem{11488474}
Z.~Ding, R.~Schober, and H.~V. Poor, ``Environment division multiple access
  ({EDMA}): A feasibility study via pinching antennas,'' \emph{IEEE Trans.
  Wireless Commun.}, vol.~25, pp. 15\,675--15\,691, 2026.

\bibitem{irs1}
M.~D. Renzo, M.~Debbah, D.-T. Phan-Huy, A.~Zappone, M.-S. Alouini, C.~Yuen,
  V.~Sciancalepore, G.~C. Alexandropoulos, J.~Hoydis, H.~Gacanin, J.~de~Rosny,
  A.~Bounceu, G.~Lerosey, and M.~Fink, ``Smart radio environments empowered by
  {AI} reconfigurable meta-surfaces: An idea whose time has come,''
  \emph{EURASIP J. on Wirel. Com. Netw.}, vol. 129, pp. 1--20, May 2019.

\bibitem{irs2}
Q.~{Wu} and R.~{Zhang}, ``Intelligent reflecting surface enhanced wireless
  network via joint active and passive beamforming,'' \emph{IEEE Trans. Wirel.
  Commun.}, vol.~18, no.~11, pp. 5394--5409, Nov. 2019.

\bibitem{10318061}
L.~Zhu, W.~Ma, and R.~Zhang, ``Modeling and performance analysis for movable
  antenna enabled wireless communications,'' \emph{IEEE Trans. Wireless
  Commun.}, vol.~23, no.~6, pp. 6234--6250, Jun. 2024.

\bibitem{9264694}
K.-K. Wong, A.~Shojaeifard, K.-F. Tong, and Y.~Zhang, ``Fluid antenna
  systems,'' \emph{IEEE Trans. Wireless Commun.}, vol.~20, no.~3, pp.
  1950--1962, Mar. 2021.

\bibitem{pinching_antenna2}
A.~Fukuda, H.~Yamamoto, H.~Okazaki, Y.~Suzuki, and K.~Kawai, ``Pinching antenna
  - using a dielectric waveguide as an antenna,'' \emph{NTT DOCOMO Technical
  J.}, vol.~23, no.~3, pp. 5--12, Jan. 2022.

\bibitem{mypa}
Z.~Ding, R.~Schober, and H.~V. Poor, ``Flexible-antenna systems: A
  pinching-antenna perspective,'' \emph{IEEE Trans. Commun.}, vol.~73, no.~10,
  pp. 9236--9253, Oct. 2025.

\bibitem{9737357}
F.~Liu, Y.~Cui, C.~Masouros, J.~Xu, T.~X. Han, Y.~C. Eldar, and S.~Buzzi,
  ``Integrated sensing and communications: Toward dual-functional wireless
  networks for {6G} and beyond,'' \emph{IEEE J. Sel. Areas Commun.}, vol.~40,
  no.~6, pp. 1728--1767, 2022.

\bibitem{9373011}
Y.~Zeng and X.~Xu, ``Toward environment-aware {6G} communications via channel
  knowledge map,'' \emph{IEEE Wireless Commun.}, vol.~28, no.~3, pp. 84--91,
  2021.

\bibitem{10955337}
Y.~Jiang, X.~Li, G.~Zhu, H.~Li, J.~Deng, K.~Han, C.~Shen, Q.~Shi, and R.~Zhang,
  ``Integrated sensing and communication for low altitude economy:
  Opportunities and challenges,'' \emph{IEEE Commun. Mag.}, vol.~63, no.~12,
  pp. 72--78, Dec. 2025.

\bibitem{CMCC}
``{C-RAN}: The road towards green {RAN},'' China Mobile Res. Inst., Beijing,
  China, Oct. 2011, White Paper, ver. 2.5.

\bibitem{Dingpoor1311}
Z.~Ding and H.~V. Poor, ``The use of spatially random base stations in cloud
  radio access networks,'' \emph{IEEE Signal Process. Lett.}, vol.~20, no.~11,
  pp. 1138--1141, Nov 2013.

\bibitem{7513863}
M.~Peng, S.~Yan, K.~Zhang, and C.~Wang, ``Fog-computing-based radio access
  networks: issues and challenges,'' \emph{IEEE Network}, vol.~30, no.~4, pp.
  46--53, 2016.

\bibitem{7827017}
H.~Q. Ngo, A.~Ashikhmin, H.~Yang, E.~G. Larsson, and T.~L. Marzetta,
  ``Cell-free massive {MIMO} versus small cells,'' \emph{IEEE Transactions on
  Wireless Communications}, vol.~16, no.~3, pp. 1834--1850, 2017.

\bibitem{11434944}
Y.~Xu, J.~Cui, Y.~Zhu, Z.~Ding, T.-H. Chang, R.~Schober, V.~W.~S. Wong, O.~A.
  Dobre, G.~K. Karagiannidis, H.~V. Poor, and X.~You, ``Generalized
  pinching-antenna systems: A tutorial on principles, design strategies, and
  future directions,'' \emph{IEEE Commun. Surveys Tuts.}, vol.~28, pp.
  5872--5908, 2026.

\bibitem{9570143}
X.~Mu, Y.~Liu, L.~Guo, J.~Lin, and R.~Schober, ``Simultaneously transmitting
  and reflecting ({STAR}) {RIS} aided wireless communications,'' \emph{IEEE
  Trans. Wireless Commun.}, vol.~21, no.~5, pp. 3083--3098, May 2022.

\bibitem{11106497}
F.~I.~G. Carvalho, R.~V. d.~O. Paiva, T.~F. Maciel, V.~F. Monteiro, F.~R.~M.
  Lima, D.~C. Moreira, D.~A. Sousa, B.~Makki, M.~Åström, and L.~Bao,
  ``Network-controlled repeater - an introduction,'' \emph{IEEE Commun.
  Standards Mag.}, vol.~10, no.~1, pp. 186--193, 2026.

\bibitem{10599123}
J.~Wang, H.~Du, D.~Niyato, J.~Kang, S.~Cui, X.~Shen, and P.~Zhang, ``Generative
  {AI} for integrated sensing and communication: Insights from the physical
  layer perspective,'' \emph{IEEE Wireless Commun.}, vol.~31, no.~5, pp.
  246--255, 2024.

\bibitem{8808168}
K.~B. Letaief, W.~Chen, Y.~Shi, J.~Zhang, and Y.-J.~A. Zhang, ``The roadmap to
  {6G}: {AI} empowered wireless networks,'' \emph{IEEE Commun. Mag.}, vol.~57,
  no.~8, pp. 84--90, 2019.

\bibitem{8714026}
N.~C. {Luong}, D.~T. {Hoang}, S.~{Gong}, D.~{Niyato}, P.~{Wang}, Y.~{Liang},
  and D.~I. {Kim}, ``Applications of deep reinforcement learning in
  communications and networking: A survey,'' \emph{IEEE Commun. Surveys Tuts.},
  vol.~21, no.~4, pp. 3133--3174, Fourth quarter 2019.

\bibitem{9448276}
Z.~Ding, R.~Schober, and H.~V. Poor, ``No-pain no-gain: {DRL} assisted
  optimization in energy-constrained {CR-NOMA} networks,'' \emph{IEEE Trans.
  Commun.}, vol.~69, no.~9, pp. 5917--5932, 2021.

\end{thebibliography}
  \end{document}